\documentclass[onecolumn,10pt,showpacs]{revtex4}
\usepackage{graphicx}
\usepackage{epsfig}
\usepackage{bm}
\usepackage{amsfonts}

\input epsf

\begin{document}

\title{\Large Dynamical System Analysis of Interacting Variable Modified Chaplygin Gas Model in FRW Universe}

\author{\bf Jhumpa Bhadra$^1$\footnote{bhadra.jhumpa@gmail.com}
and Ujjal Debnath$^1$\footnote{ujjaldebnath@yahoo.com} }

\affiliation{$^1$Department of Mathematics, Bengal Engineering and
Science University, Shibpur, Howrah-711 103, India. }

\date{\today}

\begin{abstract}
In this work, we have considered interacting dynamical model
taking variable modified Chaplygin gas which plays as dark energy
coupled to cold dark matter in the flat FRW universe. Since the
nature of dark energy and dark matter is still unknown, it is
possible to have interaction between them and we choose the
interaction term in phenomenologically. We have converted all the
equations in the dynamical system of equations by considering the
dimensionless parameters and seen the evolution of the
corresponding autonomous system. The feasible critical point has
been found and for the stability of the dynamical system about the
critical point, we linearize the governing equation around the
critical point. We found that there exists a stable scaling
(attractor) solution at late times of the Universe and found some
physical range of $n$ and the interaction parameter $c$. We have
shown that for our calculated physical range of the parameters,
the Universe explores upto quintessence stage. The deceleration
parameter, statefinder parameters, Hubble parameter and the scale
factor have been calculated around the critical point. Finally
some consequences around the critical point i.e., the distance
measurement of the Universe like lookback time, luminosity
distance, proper distance, angular diameter distance, comoving
volume, distance modulus and probability of intersecting objects
have been analyzed before and after the present age of the
Universe.
\end{abstract}


\maketitle

\section{Introduction}

Recent observations of redshift and luminosity of type Ia
supernovae \cite{{Bachall},{Perlmutter1},{Perlmutter2},{Riess}},
WMAP \cite{Bennett}, Chandra X-ray Observatory \cite{Allen} etc.
indicate that universe is spatially flat and undergoing
accelerated expansion. The most important discovery over the last
few decays is to search for the existence of dark energy which
violates the strong energy condition i.e., $\rho + 3p < 0$
\cite{{Sahni1},{Peebles0},{Padmanabhan}}. The mystifying fluid
namely dark energy is understood to dominate the 70\% of the
Universe and have enough negative pressure to drive current
acceleration 30\% dark matter (cold dark matters plus baryons).
There are various candidates to play the role of the dark energy.
The most obvious candidate for dark energy is the cosmological
constant with the equation of state $ p/\rho = -1$. The type of
dark energy represented by a scalar field is often called
quintessence. Other dark energy candidates are namely tachyonic
field \cite{Sen}, DBI-essence \cite{Martin}, Chaplygin gas
\cite{Kamenshchik}, phantom, holographic dark energy hessence dark
energy \cite{Zhang}, k-essenece \cite{Armendariz-Picon}
and dilaton dark energy \cite{Lu}.\\

Recently the  Chaplygin gas, also named quartessence,
characterized by an exotic equation of state $p=-B/\rho$ $(B>0)$,
was suggested as a candidate of a unified model of dark energy
\cite{Kamenshchik}. The Chaplygin gas behaves as pressureless
fluid for small values of the scale factor and for large values of
the scale factor as a cosmological constant, tends to accelerate
the expansion. The above equation was generalized to the form
$p=-B/\rho^\alpha$, $0\leq \alpha \leq 1$
\cite{{Gorini},{Alam},{Bento}}. Consequently this was modified to
$p=A\rho-B/\rho^\alpha$,  $0\leq \alpha \leq 1,~~ A>0$ which is
known as {\it modified Chaplygin gas}
\cite{{Benaoum},{Sahni},{Debnath}} shows a radiation era $(A =
1/3)$ when the scale factor $a(t)$ is vanishingly small and a
$\Lambda$CDM model when the scale factor $a(t)$ is infinitely
large. {\it Variable Chaplygin gas} first proposed  by Guo and
Zhang \cite{Guo} with equation of state $p=-B/\rho$, where $B$ is
a positive function of the cosmological scale factor $`a$' i.e.,
$B = B(a)$. This assumption is reasonable since $B(a)$ is related
to the scalar potential if we take the Chaplygin gas as a
Born-Infeld scalar field \cite{Bento2}. Afterward there are some
works on variable Chaplygin gas model \cite{{Sethi},{Guo2}}.
Further, Debnath \cite{Debnath1} introduced {\it variable modified
Chaplygin gas} (VMCG) for acceleration of the universe. Several
interesting features and physical interpretations of VMCG have
been shown by several authors \cite{Jamil,Chatto,Chatto1,Xing}.
The dynamical system analysis of pure and generalized Chaplygin
gas model and the nature of critical points have been analyzed for
Einstein's gravity and loop quantum gravity \cite{Wu, Wu1, Zhang1, Jamil1, Jamil2, Jamil3}.\\

In this work, we consider a model of interacting variable modified
Chaplygin gas (VMCG) with dark matter in the framework of Einstein
gravity. We construct the formalism of autonomous dynamical system
of equations for this interacting model. Since the nature of dark
energy and dark matter is still unknown, it is possible to have
interaction between them and we choose the interaction term in
phenomenologically. We convert them to dimensionless form and
perform stability analysis and solve them numerically. We obtain a
stable scaling solution (which is also an `attractor') of the
equations in FRW model. Some consequences like lookback time,
proper distance, luminosity distance, angular diameter distance,
comoving volume, distance modulus and probability of intersecting
objects of the solution around the critical point have been
investigated. We discuss our results in the final section.

\section{Dynamical model of interacting VMCG and Dark matter }

We consider a spatially flat universe with  VMCG as dark energy
and dark matter interacting through an interaction term. Thus the
Einstein equations and continuity equation of VMCG and dark matter
can be written respectively as (choosing $8\pi G=c=1$)

\begin{eqnarray} H^2 = \frac{1}{3} \rho \end{eqnarray}
\begin{eqnarray} \dot{H}=-\frac{1}{2}\left(p + \rho \right)\end{eqnarray}
and
\begin{eqnarray} \dot{\rho}+3H (\rho+p)=0 \end{eqnarray}

where $\rho = \rho_{vmcg}+\rho_{dm}$ and $p = p_{vmcg}+p_{dm}$
($p_{dm}$ is very small quantity, somewhere dark matter assumed as
pressureless quantity) are the total cosmic energy density and
pressure respectively with the subscripts $vmcg$ and $dm$ denote
the VMCG and dark matter respectively. Since we consider that the
VMCG and dark matter do not conserve separately, so their
continuity equations are

\begin{eqnarray}\dot{\rho}_{vmcg}+3H \left(\rho_{vmcg}+p_{vmcg} \right)=-Q \end{eqnarray}
and
\begin{eqnarray}\dot{\rho}_{dm}+3H \left(\rho_{dm}+p_{dm} \right)=Q\end{eqnarray}

To obtain a suitable evolution of the Universe an interaction is
often assumed such that the decay rate should be proportional to
the present value of the Hubble parameter for good fit to the
expansion history of the Universe as determined by the Supernovae
and CMB data \cite{Berger}. Here $Q$ is the interaction term which
have dimension of density multiplied by Hubble parameter. For
suitable choice of $Q$ is $Q=3cH\rho$, $c$ is the coupling
parameter denoting the transfer strength. The interaction term can
not be trace out from the first principles due to unidentified
scenario of both the dark energy and dark matter. If $Q<0$ then
energy density of dark energy become negative at sufficiently
early times, therefore the second law of thermodynamics can be
violated \cite{Alcaniz}. Thus $Q$ must be positive and small. Also
the observational data of 182 Gold type Ia supernova samples, CMB
data from the three year WMAP survey and the baryonic acoustic
oscillations from the Sloan Digital Sky Survey estimated that the
coupling parameter between dark matter and dark energy must be a
small positive value (of the order unity), which satisfies the
requirement for solving the cosmic
coincidence problem and the second law of thermodynamics \cite{Feng}. \\

The VMCG equation of state is given by \cite{Debnath1}

\begin{eqnarray}
p_{vmcg}=A \rho_{vmcg} -\frac{B(a)}{\rho_{vmcg}^\alpha},~~ \mbox{
with $A$, $\alpha$ are constants and $0\leq\alpha\leq1$}
\end{eqnarray}

where $B(a)$ is a positive function of the cosmological scale
factor $`a$'. Now consider for simplicity, $B(a)=B_0 a^{-n}$,
where $n$ and $B_0$ are constants. More precisely
the restriction for accelerating universe becomes $0< n\leq 4$ \cite{Debnath1}.\\

The dark matter equation of state is
\begin{eqnarray}p_{dm}=w_{dm} \rho_{dm}, \mbox{ with $w_{dm}$ is a small constant.}\end{eqnarray}

To analyze the dynamical system, we convert the physical parameter
into dimensionless form as follows:
\begin{eqnarray} x=\ln a,~~ u=\Omega_{vmcg}=\frac{\rho_{vmcg}}{3H^2} \mbox{~~and~~} v=\frac{p_{vmcg}}{3H^2}\end{eqnarray}

The equation of state of the VMCG can be expressed as
\begin{eqnarray} w_{vmcg}(x) =\frac{p_{vmcg}}{\rho_{vmcg}}= \frac{v}{u} \end{eqnarray}

Defining dimensionless density parameters of dark matter
$\Omega_{dm}=\frac{\rho_{dm}}{3H^2}$ and using the Friedmann
equation (1), we obtain
\begin{eqnarray} \Omega_{dm}=1-\Omega_{vmcg}=1-u\end{eqnarray}

Thus for the flat universe, $u$ must lies in the region $0\leq
u\leq 1$, since the energies of VMCG and dark matter can not be negative. \\

Now we can cast the evolution equations in the following
autonomous system of $u$ and $v$ in the form:

\begin{eqnarray} \frac{du}{dx}=-3c-3\left(u-1\right)\left(u w_{dm}-v \right)~,\end{eqnarray}
\begin{eqnarray} \frac{dv}{dx}=\left\{A(1+\alpha)-\alpha \frac{v}{u}\right\}\left\{-3c -3\left(1+\frac{v}{u}\right)u\right\}
+n\left(A-\frac{v}{u}\right)u \nonumber\\
+3v\left\{\left(1+\frac{v}{u}\right)u+\left(1+w_{dm}\right)\left(1-u
\right)\right\}\end{eqnarray}

\subsection{Critical point}

The critical points of the above system are the solution of the
equations $\frac{du}{dx}=\frac{dv}{dx}=0$. The only feasible
critical point is obtained as

\begin{eqnarray}u_{crit}=\frac{n-3\left(1-c+w_{dm}\right)\left(1+\alpha\right)}{n-3\left(1+w_{dm}\right)\left(1+\alpha\right)}~~,~~~~~~~~~~~~
~~~~~~~~~~~~~~~~~~~~~~~~\\
v_{crit}=-\frac{n^2-3n\left(2+w_{dm}\right)\left(1+\alpha\right)+9\left(1+w_{dm}+cw_{dm}\right)\left(1+\alpha\right)^2}
{3\left(1+\alpha\right)\left\{3\left(1+w_{dm}\right)\left(1+\alpha\right)-n\right\}}\end{eqnarray}

Since in a spatially flat universe, the physically meaningful
range of $u$ is $0 \leq u \leq 1$, hence $0 \leq u_{crit} \leq 1$
leads to the condition $0<c\leq\frac{3\left(1+w_{dm}\right)
\left(1+\alpha\right)-n}{3 \left(1+\alpha\right)}$ along with
$n\leq min\{3(1+w_{dm})(1+\alpha),4\}$, \cite{Debnath1} for
existence of the critical point. This condition of $c$ represents
there is an energy transfer from to dark matter to VMCG.\\

\subsection{Stability around critical point}

For the stability of the dynamical system about the critical
point, we linearize the governing equation around the critical
point i.e., $u=u_{crit}+\delta u$ and $v=v_{crit}+\delta v$, we
obtain

\begin{eqnarray}\delta \left(\frac{du}{dx}\right)=\left[3\left(v+w_{dm}-2u w_{dm}\right)\right]_{crit}
\delta u+\left[3\left(-1+u\right)\right]_{crit} \delta v,\\
\delta
\left(\frac{dv}{dx}\right)=\left[-3w_{dm} v-\frac{3v(c+v)\alpha}{u^2}+A\{n-3(1+\alpha)\} \right]_{crit} \delta u+\nonumber\\
\left[\{3-3A-n+6v+3(1-u)w_{dm}\}+\frac{3(c+u-Au+2v)\alpha}{u}\right]_{crit} \delta v\end{eqnarray}\\

We find the eigen values of the Jacobian matrix at the critical
point (for simplicity we set $\alpha=1$) as in the following:

\begin{eqnarray}\lambda_{1,2}=\frac{1}{4}\left[3n-6(3+2A+c+w_{dm})+\frac{36c^2}{6c+n-6(1+w_{dm})}\right]\pm~~~~\nonumber\\
\frac{1}{4}\sqrt{\frac{\left\{(6-n)^2+72cw_{dm}-36w_{dm}^2+12A\left(6-6c-n+6w_{dm}\right)\right\}}{\{6c+n-6(1+w_{dm})\}}}\times \nonumber\\
\sqrt{\frac{\left\{-144c^2+(6-n)^2+12A\left(6-6c-n+6w_{dm}\right)-36w_{dm}^2-24c\left(n-6-9w_{dm}\right)\right\}}{\{6c+n-6(1+w_{dm})\}}}
\end{eqnarray}

\vspace{1in}
\begin{figure}[!h]

\epsfxsize = 3.0 in \epsfysize = 2 in
\epsfbox{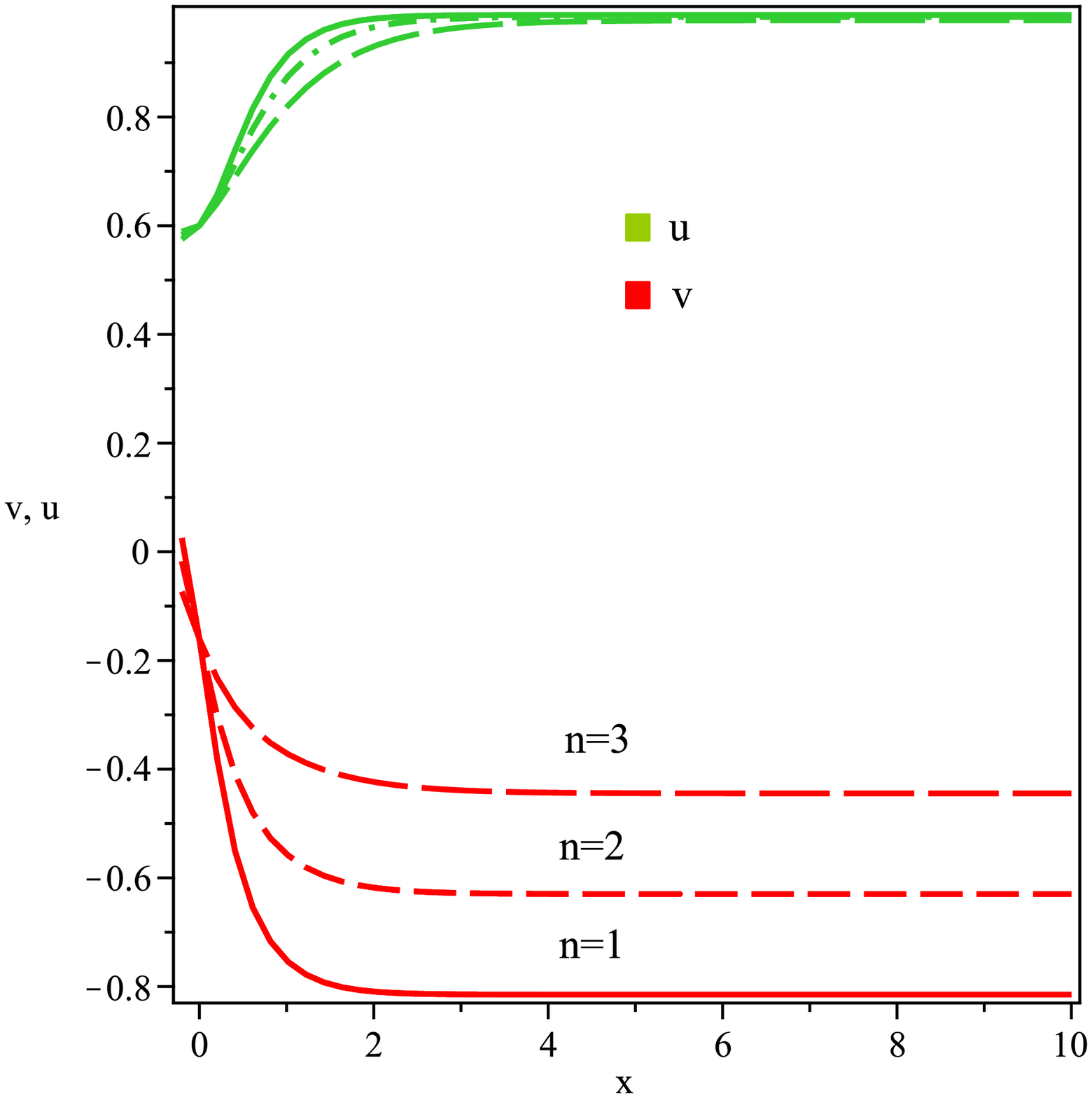}~~~\epsfxsize = 3.0 in \epsfysize = 2 in
\epsfbox{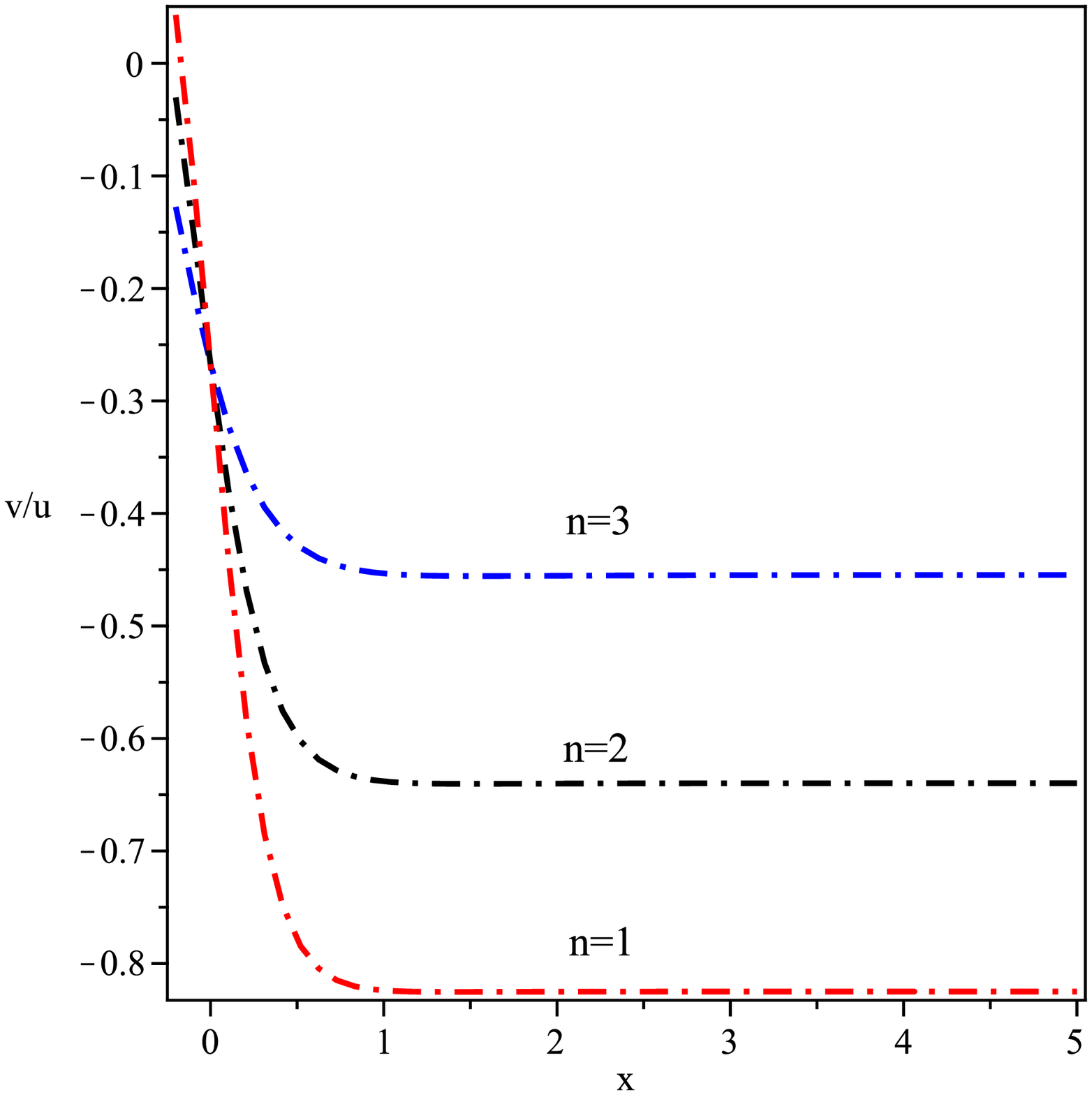}\\
~~FIG.1~~~~~~~~~~~~~~~~~~~~~~~~~~~~~~~~~~~~~~~~~~~~~~~~~~~~~~~~~~~~~~~~~~~~FIG.2\\

\caption{The dimensionless parameters are plotted against
e-folding time for different values of $n$. The initial condition
is $v(0) = -0.16, u(0) = 0.6$. Other parameters are fixed at $c =
0.01, A = 0.3$, $w_{dm}=0.01$ and $\alpha$ = 0.8.}

\caption{The evolution of $w(x)=v/u$ for the interacting VMCG
model, corresponding to the initial conditions $u(0) = 0.6$, $v(0)
= -0.016$; The figure is drawn for different values of $n$, other
parameters are fixed at $c = 0.01, A = 0.3$, $w_{dm}=0.01$ and
$\alpha$ = 0.8 respectively.}
\end{figure}

If the real parts of the above eigenvalues are negative, the
critical point is stable node and is a stationary attractor
solution; otherwise unstable and thus oscillatory. Physical
meaningful range of $c$ is $0 \leq c\leq
\frac{3\left(1+w_{dm}\right)\left(1+\alpha\right)-n}{3
\left(1+\alpha\right)}$ and in this range the critical point
$(u_{crit},v_{crit})$ is stable and is a late-time stationary
attractor solution. Here we plot some figures to show the
properties of evolution of the Universe controlled by the
dynamical system (11) and (12). The dimensionless parameters $u$
and $v$ have been drawn in figure 1 in terms of $x=\ln a$. Also
the EOS parameter $w_{vmcg}=\frac{v}{u}$ is drawn in figure 2 for
$n=1,2,3$. We see that $u$ becomes positive but $v$ and
$\frac{v}{u}$ explore to the negative level above $-1$. The
phase-space diagrams have been shown in figures 3 and 4 for
$n=1,3$ respectively. We see that from the progressions of the
phase-portrait, $u$ goes to $1$ and $v$ tends to a negative value
above $-1$ and hence the solution is stationary attractor and thee
corresponding critical point is a stable node.

\vspace{1in}
\begin{figure}[!h]

\epsfxsize = 3.0 in \epsfysize =2 in
\epsfbox{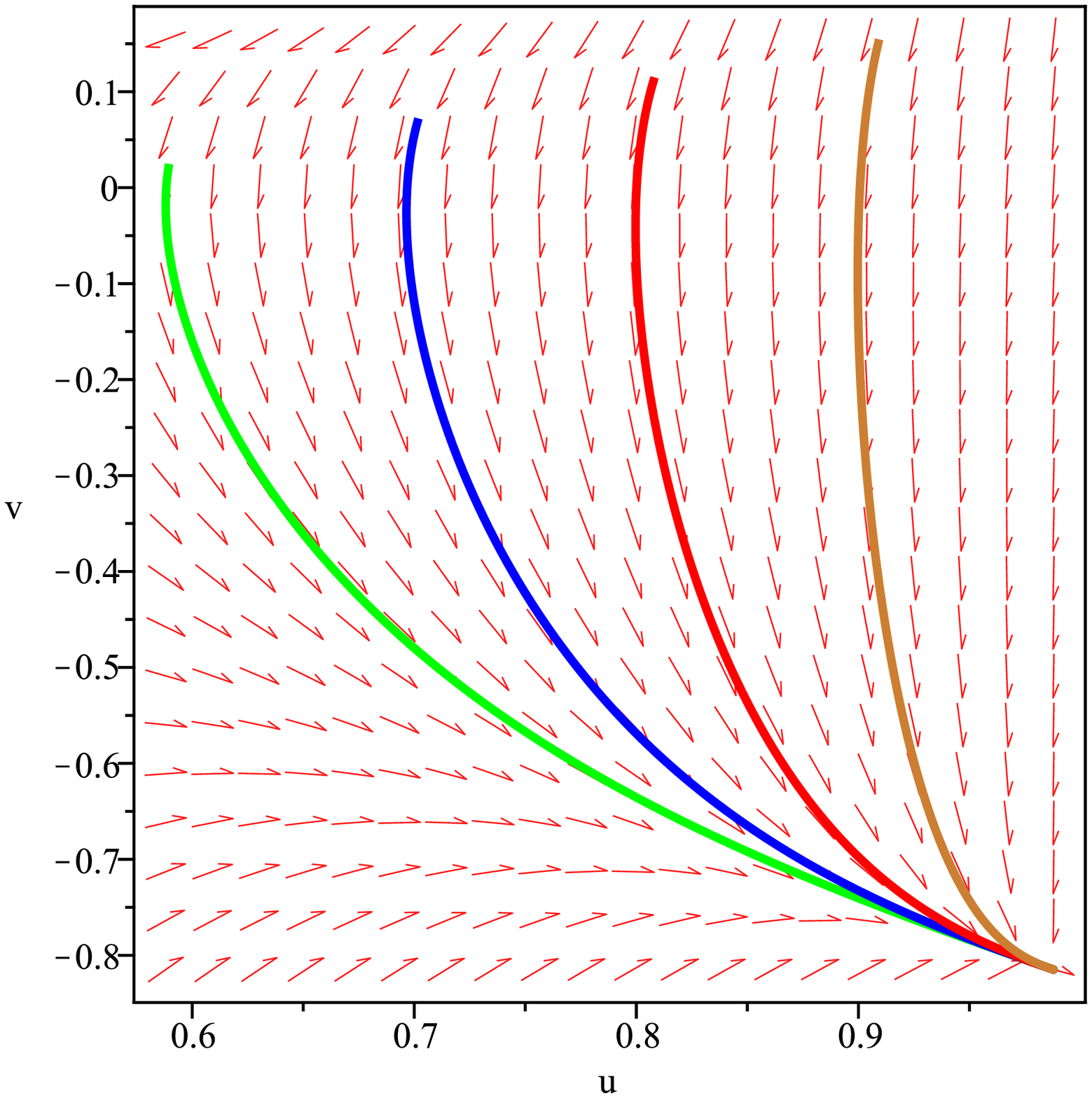}~~~~~~\epsfxsize = 3.0 in \epsfysize =2 in
\epsfbox{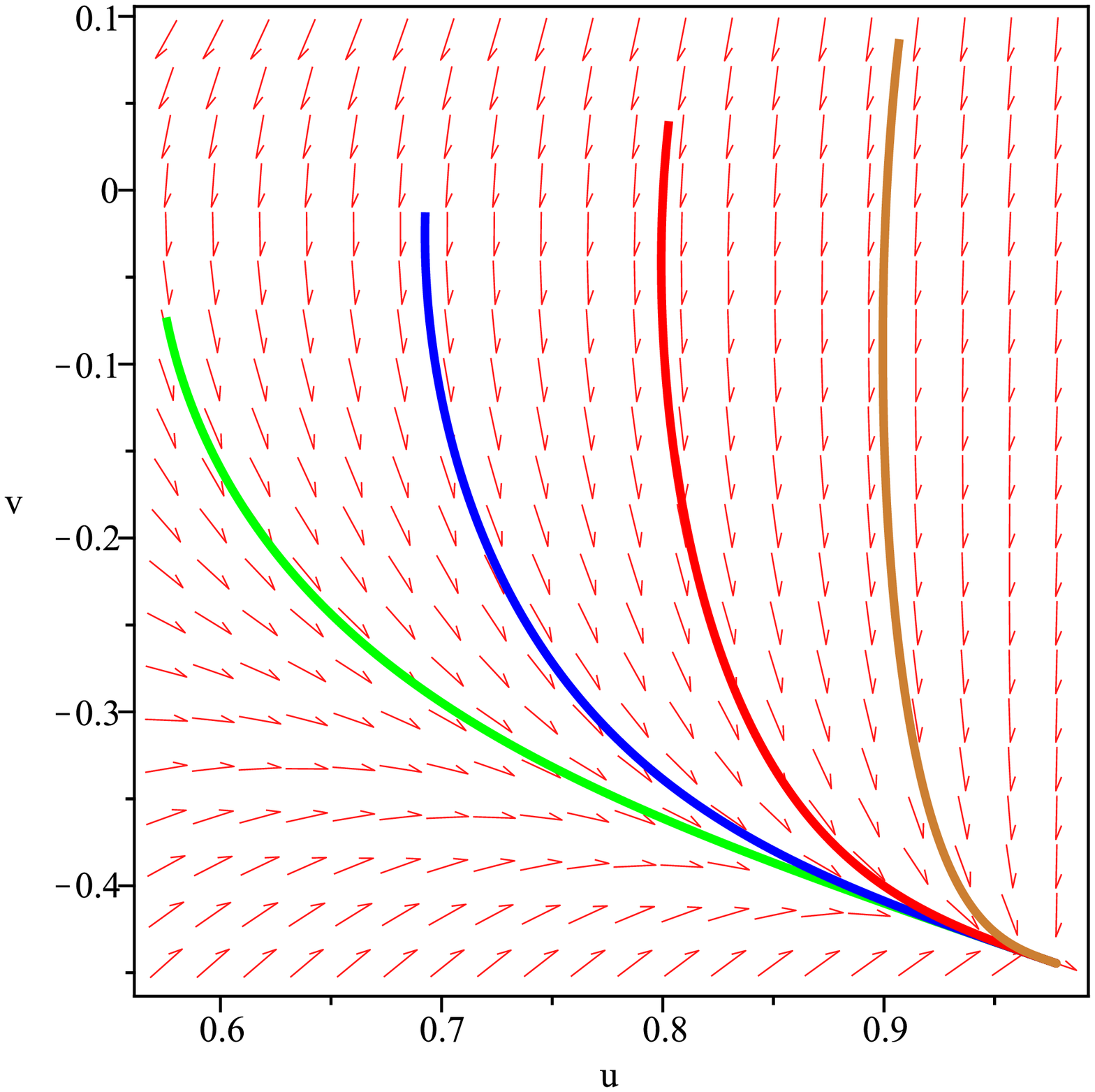}\\

~~FIG.3~~~~~~~~~~~~~~~~~~~~~~~~~~~~~~~~~~~~~~~~~~~~~~~~~~~~~~~~~~~~~~~~~~~~FIG.4\\

\caption{The phase space diagram of parameters depicting an
attractor solution. The initial conditions chosen are $v(0) =
-0.16$, $u(0) = 0.6$ (green); $v(0) =- 0.12$, $u(0) = 0.6$ (blue);
$v(0) =- 0.08$, $u(0) = 0.7$ (red); $v(0) =- 0.4$, $u(0) = 0.8$
(brown). Other parameters are fixed at $c = 0.01$, $A = 0.3$, $w_{dm}=0.01$, $\alpha =0.8$, $n=1$.}

\caption{The phase space diagram of parameters depicting an
attractor solution. The initial conditions chosen are $v(0) =
-0.16$, $u(0) = 0.6$ (green); $v(0) =- 0.12$, $u(0) = 0.6$ (blue);
$v(0) =- 0.08$, $u(0) = 0.7$ (red); $v(0) =- 0.4$, $u(0) = 0.8$
(brown). Other parameters are fixed at $c = 0.01$, $A = 0.3$,
$w_{dm}=0.01$, $\alpha =0.8$, $n=3$.}
\end{figure}

\vspace{1in}
\begin{figure}[!h]

\epsfxsize = 3.0 in \epsfysize =2 in
\epsfbox{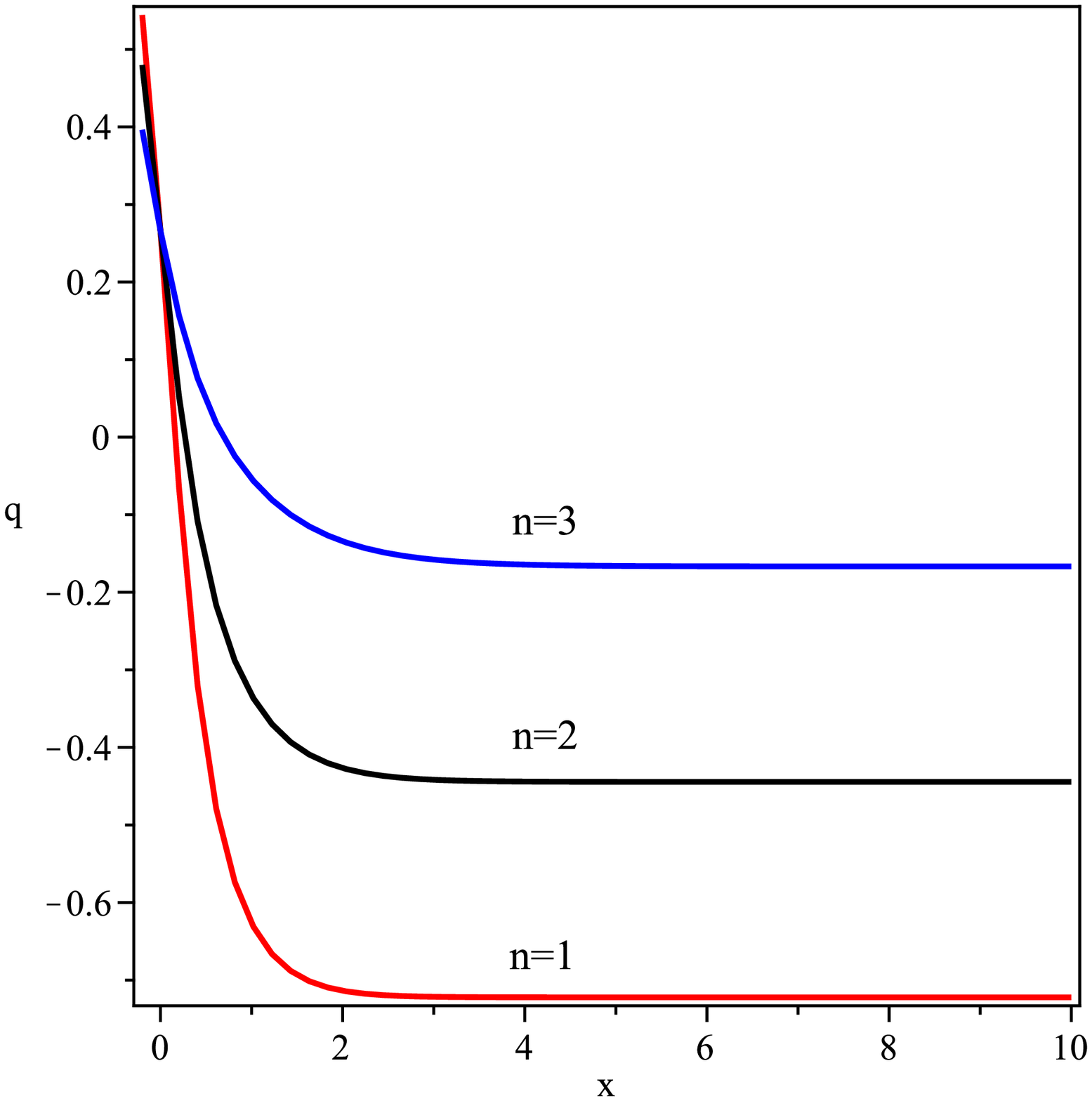}~~~~~~\epsfxsize = 3.0 in \epsfysize =2 in
\epsfbox{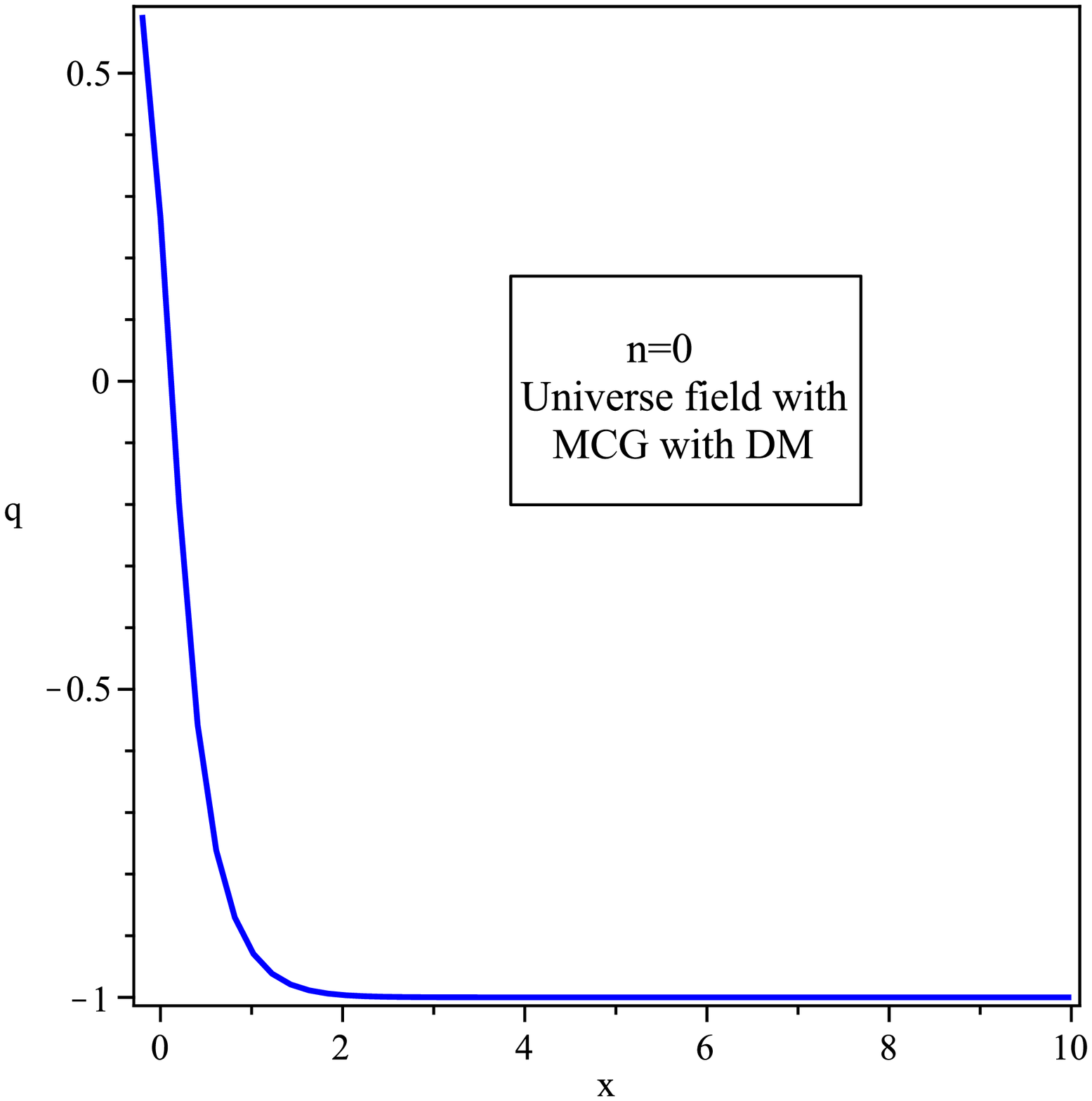}\\

~~FIG.5~~~~~~~~~~~~~~~~~~~~~~~~~~~~~~~~~~~~~~~~~~~~~~~~~~~~~~~~~~~~~~~~~~~~FIG.6\\

\caption{The deceleration parameter $q$ is plotted against the
evolution of the universe for different values of $n$. Other
parameters are fixed at $c = 0.01$, $A = 0.3$,
$w_{dm}=0.01$, and $\alpha =0.8$.}

\caption{The deceleration parameter $q$ is plotted against the
evolution of the universe for $n=0$. Other parameters are fixed at
$c = 0.01$, $A = 0.3$, $w_{dm}=0.01$, and $\alpha =0.8$.}
\end{figure}

At the critical point, the deceleration parameter
$q=-1-\left(\dot{H}/H^2\right)$ is obtained as
\begin{eqnarray}q=-1+\frac{3}{2}X,~~~~ X=1+v_{crit}+w_{dm}(1-u_{crit})\end{eqnarray}

Integrating, we have the Hubble parameter as (ignoring the integrating constant)
\begin{eqnarray}H=\frac{2}{3Xt}\end{eqnarray}

Again integrating $(19)$, we get

\begin{eqnarray}a(t)=a_0 \left(\frac{t}{t_0}\right)^{\frac{2}{3X}}\end{eqnarray}

which gives power law  expansion of the
universe, $a_0$ is the present value of the scale factor when
$t=t_0$ is the present time which is given by
$t_0=\frac{2}{3XH_0}$. Since we have obtained the deceleration
parameter $q$, Hubble parameter $H$ and scale factor $a(t)$ around
the critical point, so these are valid for the late stage of the
evolution of the universe i.e., when time $t$ is very large. Thus
$X$ should be taken positive value for positivity of the Hubble
parameter $H$ and hence the scale factor $a(t)$ is increasing
function of time $t$. So from (18), we obtain $q\in(-1,0)$ for
accelerating phase of the universe if $X\in(0,2/3)$. So the VMCG
couples with dark matter filled in the universe always drives
acceleration. Figure 5 shows the graph of $q$ against $x=\ln a$
for some particular values of $n=1,2,3$. For $n=0$, the VMCG model
reduces to MCG and these couples with dark matter filled in the
universe always drives acceleration and this model explores upto
$\Lambda$CDM stage ($q=-1$) (figure 6).

\subsection{Statefinder parameters}

We also determine the dimensionless pair of cosmological
diagnostic pair $\{r,s\}$ dubbed as statefinder parameters
introduced by Sahni et al \cite{Sahni}. The two parameters have a
great geometrical significance since they are derived from the
cosmic scale factor alone, though one can rewrite them in terms of
the parameters of dark energy and matter. Furthermore, the pair
characterize the properties of dark energy in a model-independent
manner i.e. independent on the theory of gravity. Also this pair
generalizes the well known geometrical parameters like the Hubble
parameter and the deceleration parameter. The parameter $r$ forms
the next step in the hierarchy of geometrical cosmological
parameters $H$ and $q$. The parameters are given by

\begin{eqnarray}r\equiv \frac{\dddot{a}}{aH^3}=\left(1-\frac{3X}{2}\right)\left(1-3X\right),~~~~s\equiv\frac{r-1}{3(q-1/2)}=X\end{eqnarray}

where $X$ is obtained in equation (18). Since we have already
obtained that for accelerating universe, $X<2/3$. In this range,
$s$ is always positive. But $r>0$ for $X<1/3$ and $r<0$ for
$1/3<X<2/3$.

\vspace{1in}
\begin{figure}[!h]

\epsfxsize = 3.0 in \epsfysize =2 in
\epsfbox{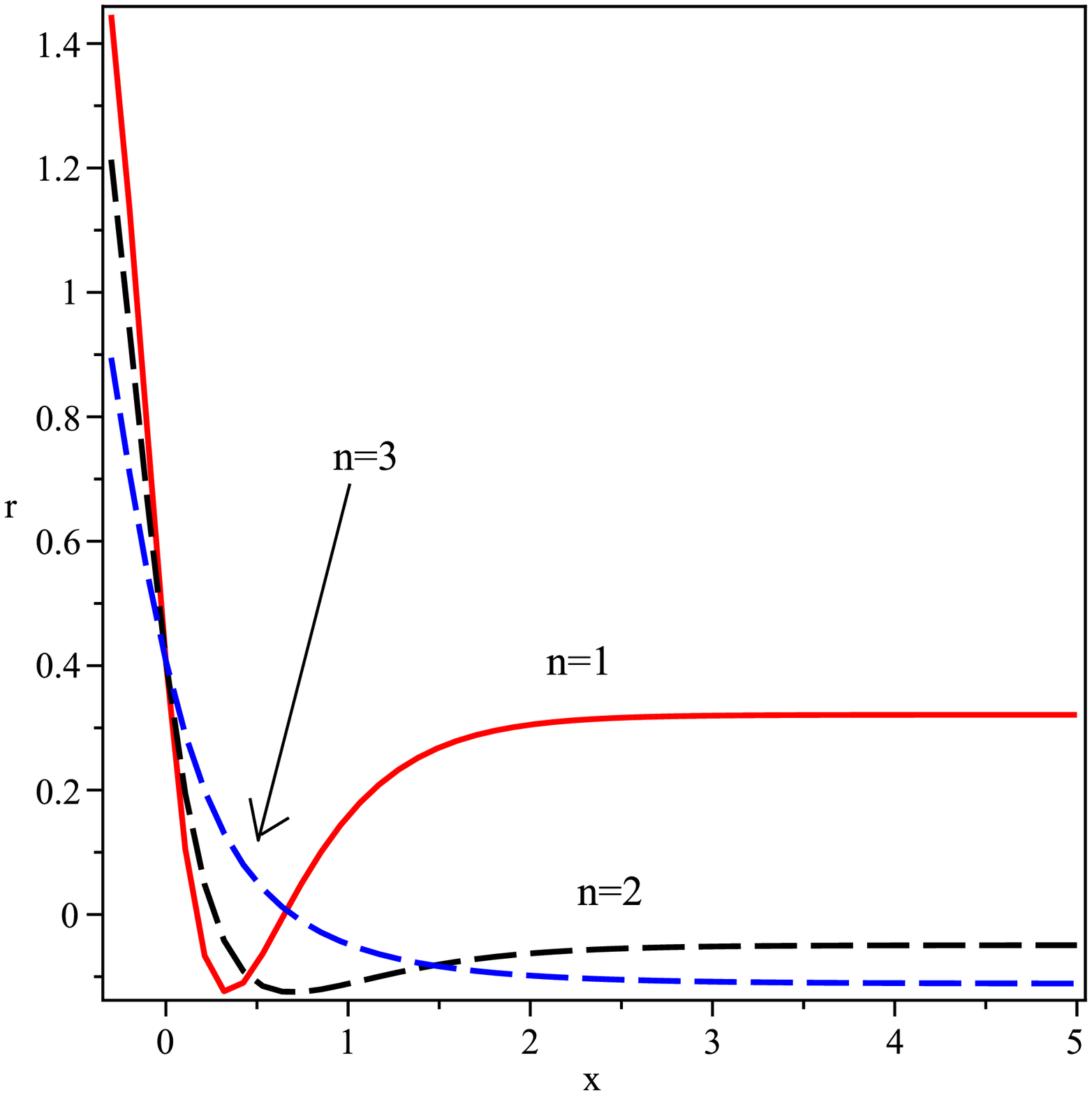}~~~~\epsfxsize = 3.0 in \epsfysize =2 in
\epsfbox{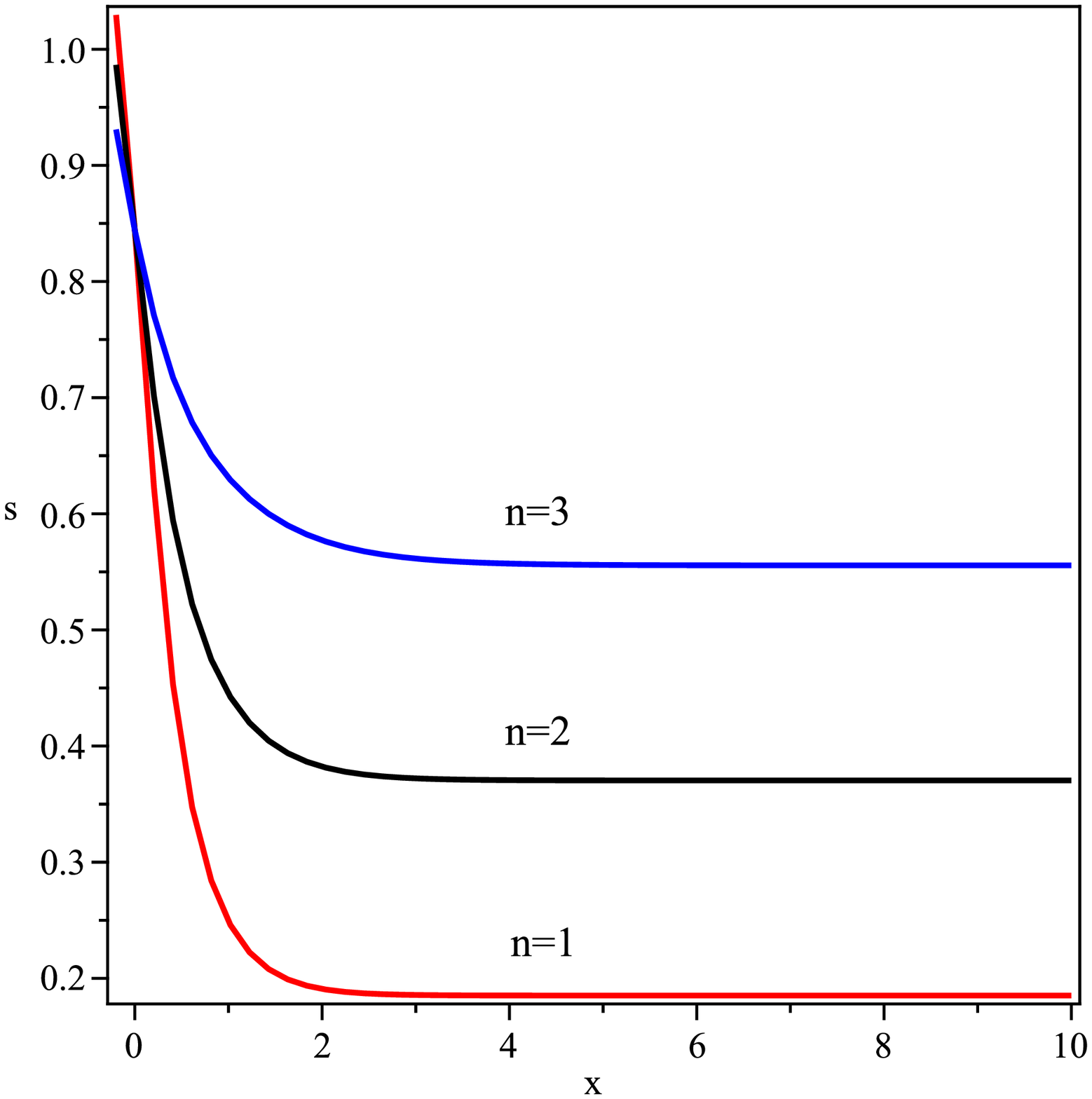}\\

~~FIG.7~~~~~~~~~~~~~~~~~~~~~~~~~~~~~~~~~~~~~~~~~~~~~~~~~~~~~~~~~~~~~~~~~~~~FIG.8\\

\caption{The statefinder parameter $r$ is plotted against the
evolution of the universe for different values of $n$. Other
parameters are fixed at $c = 0.01$, $A = 0.3$,
$w_{dm}=0.01$, and $\alpha =0.8$.}

\caption{The statefinder parameter $s$ is plotted against the
evolution of the universe for different values of $n$. Other
parameters are fixed at $c = 0.01$, $A = 0.3$, $w_{dm}=0.01$, and
$\alpha =0.8$.}
\end{figure}

\section{Consequences: Distance Measurement of the Universe}

In {\it cosmography} (the measurement of the Universe) there are
many ways to specify the distance between two points, because in
the expanding and accelerating Universe, the distances between
comoving objects are constantly changing, and Earth-bound
observers look back in time as they look out in distance. The
unifying aspect is that all distance measures somehow measure the
separation between events on radial null trajectories, i.e.,
trajectories of photons which terminate at the observer. Here we
will compute various cosmological distance measures. In this
section, we shall discuss the lookback time, luminosity distance,
proper distance, angular diameter distance, comoving volume,
distance modulus and probability of intersecting objects.

\subsection{Lookback Time}

The {\it lookback time} to an object is the difference between the
age of the Universe now (at observation) and the age of the
Universe at the time the photons were emitted (according to the
object). As light travels with finite speed, it takes time for it
to cover the distance related to the redshift it encountered. So,
a look into space is always a look back in time. It is used to
predict properties of high redshift objects with evolutionary
models, such as passive stellar evolution for galaxies. Thus if a
photon emitted by a source at the instant $t$ and received at the
time $t_0$ then the photon travel time or the lookback time $t_0
-t$ is defined by \cite{Debnath2, Arbab, Hogg, Peebles, Kolb}

\begin{eqnarray} t-t_0=\int^{a}_{a_0}{\frac{da}{\dot{a}}}\end{eqnarray}\\

where $a_0$ is the present value of the scale factor of the
universe and can be obtained from (20) at $t = t_0$. The redshift
is an important observable as they can be measured easily from the
spectral lines and the redshift increases of an object with its
distance from us. Lookback time is used to predict properties of
high-redshift objects with evolutionary models, such as passive
stellar evolution for galaxies. The redshift $z$ can be defined by
\begin{eqnarray} \frac{a_0}{a}=1+z=\left(\frac{t_0}{t}\right)^\frac{2}{3X}\end{eqnarray}\\

which gives the lookback time in the following form

\begin{eqnarray} t-t_0=\frac{2}{3XH_0}\left\{\frac{1}{\left(1+z\right)^{\frac{3X}{2}}}-1\right\}\end{eqnarray}

For accelerating universe we have already get $X<\frac{2}{3}$.
Early universe is represented by $z\rightarrow \infty$ implies
$t\rightarrow0$ and late universe $z\rightarrow -1$, which
equivalently implied $t\rightarrow \infty$. Also $z\rightarrow 0$
gives the present age $t\rightarrow t_0$ of the universe.

\subsection{Proper Distance}

As light needs time to get from an object to the observer, one can
define a distance that may be measured between the observer and
the object with a ruler at the time the light was emitted, the
{\it proper distance}. When a photon emitted by a source and
received by an observer at time $t_0$ then the proper distance
between them is defined by \cite{Debnath2, Arbab, Hogg, Peebles,
Weinberg, Weedman}

\begin{eqnarray} d=a_0 \int^{a_0}_{a}{\frac{da}{a \dot{a}}}=a_0 \int^{t_0}_{t}{\frac{dt}{a}} \end{eqnarray}

which gives
\begin{eqnarray}d=\frac{2}{H_0 (3X-2)}\left\{1-\frac{1}{\left(1+z\right)^{\frac{3X}{2}-1}}\right\}\end{eqnarray}

The proper distance may also called the {\it comoving distance}
(line of sight) of the Universe in today. So between two nearby
objects in the Universe, the distance between them which remains
constant with epoch if the two objects are moving with the Hubble
flow. In other words, it is the distance between them which would
be measured with rulers at the time they are being observed
divided by the ratio of the scale factor of the Universe then to
now. Next thing to be defined is the {\it transverse comoving
distance}, which is a quantity used to get the comoving distance
perpendicular to the line of sight. For flat universe, the
transverse comoving distance is always identical to the comoving
distance (line of sight). That means the transverse comoving
distance $=$ proper distance $= d$, for our model of the flat FRW
Universe.

\vspace{.5in}
\begin{figure}[!h]

\epsfxsize = 3.0 in \epsfysize =2 in
\epsfbox{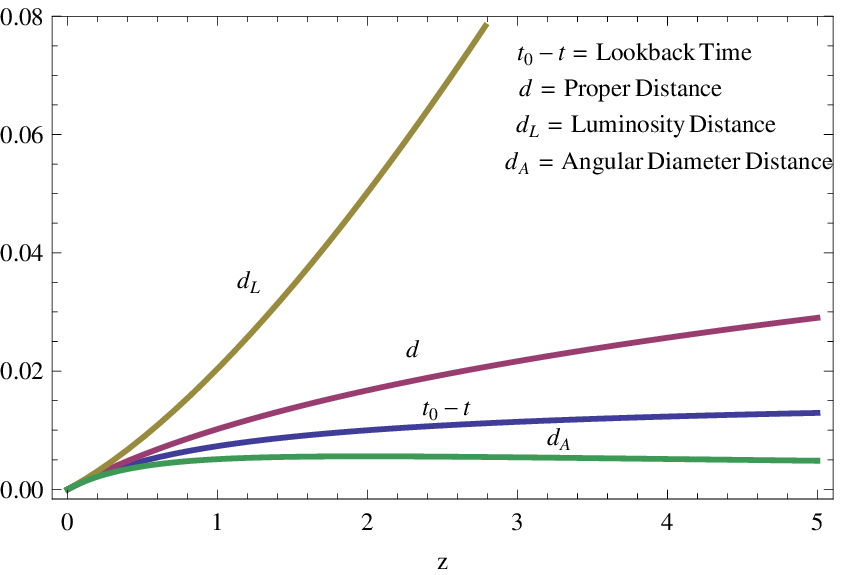}~~~~~\epsfxsize = 3.0 in \epsfysize =2 in
\epsfbox{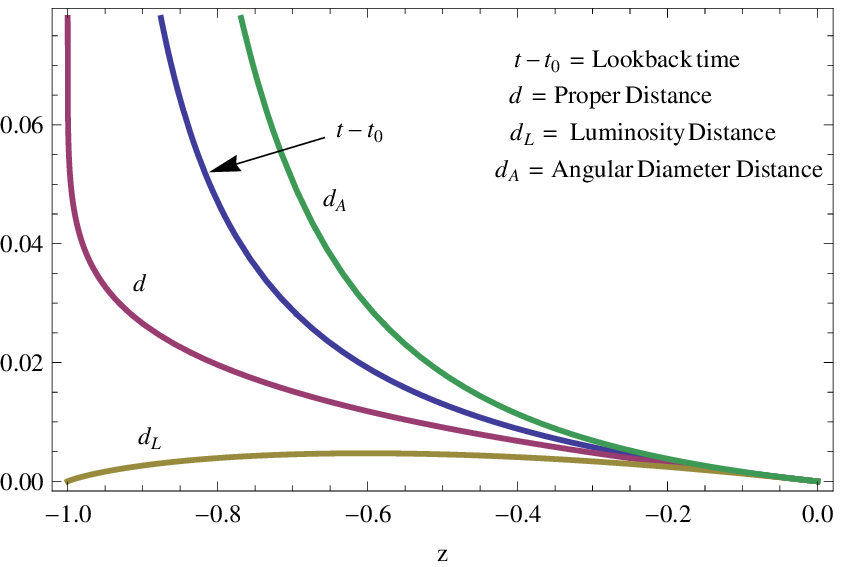}\\

~~FIG.9~~~~~~~~~~~~~~~~~~~~~~~~~~~~~~~~~~~~~~~~~~~~~~~~~~~~~~~~~~~~~~~~~~~~FIG.10\\

\caption{Look back time, proper distance, luminosity distance and
angular diameter distance as a function of redshift assuming
$H_0=72 km/s/Mpc$ before the present age of the universe.}

\caption{Look back time, proper distance, luminosity distance and
angular diameter distance as a function of redshift assuming
$H_0=72 km/s/Mpc$ after the present age of the universe.}
\end{figure}

\subsection{Luminosity Distance}

If $L$ be the total energy emitted by the source per unit time and
$\ell$ be the apparent luminosity of the object then the {\it
luminosity distance} is defined by \cite{Debnath2, Arbab, Hogg,
Weinberg, Weedman}
\begin{eqnarray}d_L=\left(\frac{L}{4 \pi \ell}\right)^{\frac{1}{2}}=d(1+z)=
\frac{2}{H_0(3X-2)}\left\{(1+z)-\frac{1}{(1+z)^{\frac{3X}{2}}}\right\}\end{eqnarray}

\subsection{Angular Diameter Distance}
The {\it angular diameter} of a light source of proper distance
$D$ observed at $t_0$ is defined by \cite{Debnath2, Arbab, Hogg,
Peebles, Weinberg, Weedman}
\begin{eqnarray}\delta=\frac{D(1+z)^2}{d_L}\end{eqnarray}

The {\it angular diameter distance} $(d_A)$ is defined as the
ratio of the source diameter to its angular diameter (in radians)
as

\begin{eqnarray}d_A=\frac{D}{\delta}=d_L(1+z)^{-2}=d(1+z)^{-1}\end{eqnarray}

For our model angular diameter distance $(d_A)$ is given by,

\begin{eqnarray}d_A=\frac{2}{H_0 (3X-2)}\left\{\frac{1}{1+z}-\frac{1}{\left(1+z\right)^{\frac{3X}{2}-2}}\right\} \end{eqnarray}

It is used to convert angular separations in telescope images into
proper separations at the source. It is famous for not increasing
indefinitely as $z\rightarrow \infty $; it turns over at $z \sim
1$ and thereafter more distant objects actually appear larger in
angular size. The angular diameter distance is maximum at

\begin{eqnarray}z_{max}=\left(\frac{2}{3X-2}\right)^{\frac{2}{3(2-X)}}-1\end{eqnarray}

and corresponding maximum angular diameter $d_A|_{max}$ taking the form

\begin{eqnarray}d_A|_{max}=\frac{1}{H_0(3X-2)}\left[2^{1+\frac{2}{3(X-2)}}\left(\frac{1}{3X-4}\right)^{\frac{2}{3(X-2)}}-
2\left\{4^{\frac{1}{3(2-X)}}\left(\frac{1}{3X-4}\right)^{\frac{2}{3(X-2)}}\right\}^{2-\frac{3X}{2}}\right]\end{eqnarray}

Look back time, proper distance, luminosity distance and angular
diameter distance before and after the present age of the universe
are respectively drawn in figures 9 and 10.

\subsection{Comoving Volume}

The {\it comoving volume} $V_C$ is the volume measure in which
number densities of non-evolving objects locked into Hubble flow
are constant with redshift as \cite{Hogg, Peebles, Weinberg}

\begin{eqnarray} dV_C= D_H \frac{(1+z)^2 d_A}{E(z)} d\Omega dz=\frac{1}{H_0} (1+z)^{2-\frac{3X}{2}}d_A^2 d\Omega dz \end{eqnarray}

where $d\Omega$ is the solid angle element and $d_A$ is the
angular diameter, $E(z)=\frac{H(z)}{H_0}$ and $D_H=\frac{c}{H_0}$
is the Hubble distance ($c$ is the velocity of light) and in our
model we assume $c=1$, $d \Omega=1$, $H_0=72 km/s/Mpc$. So, the comoving volume
is proper volume times the ratio of scale factors now to then to
the third power. Comoving volume element $dV_C/dz$ before and
after the present age of the universe are respectively drawn in
figures 11 and 12. We see that when $z>0$, the volume element
decreases as $z$ decreases and when $z<0$, the volume element
increases as $z$ decreases.

\begin{figure}[!h]

\epsfxsize = 3.0 in \epsfysize =2 in
\epsfbox{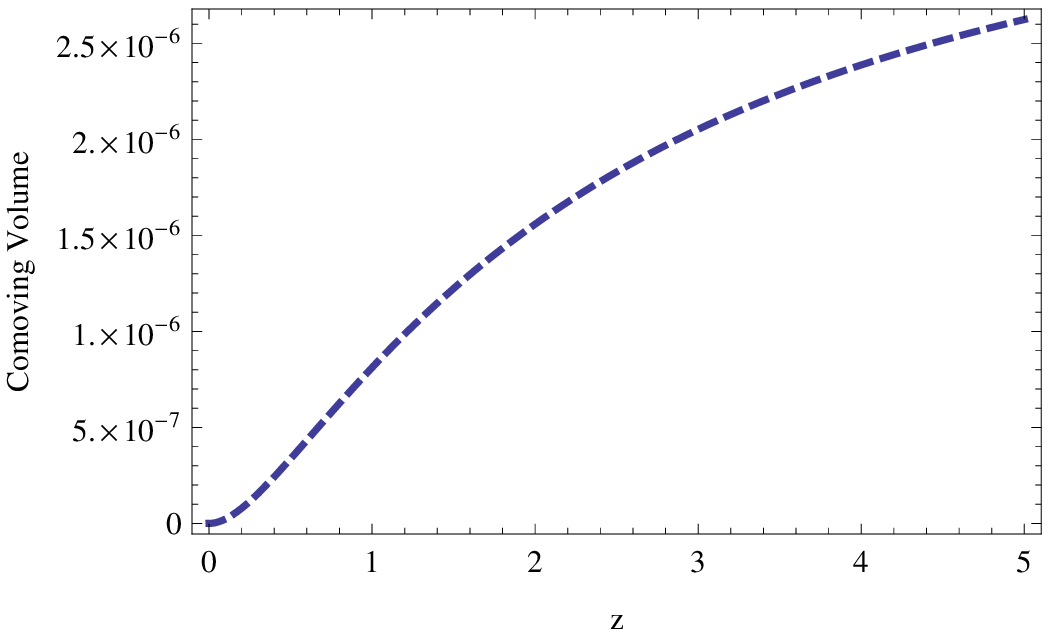}~~~~\epsfxsize = 3.0 in \epsfysize =2 in
\epsfbox{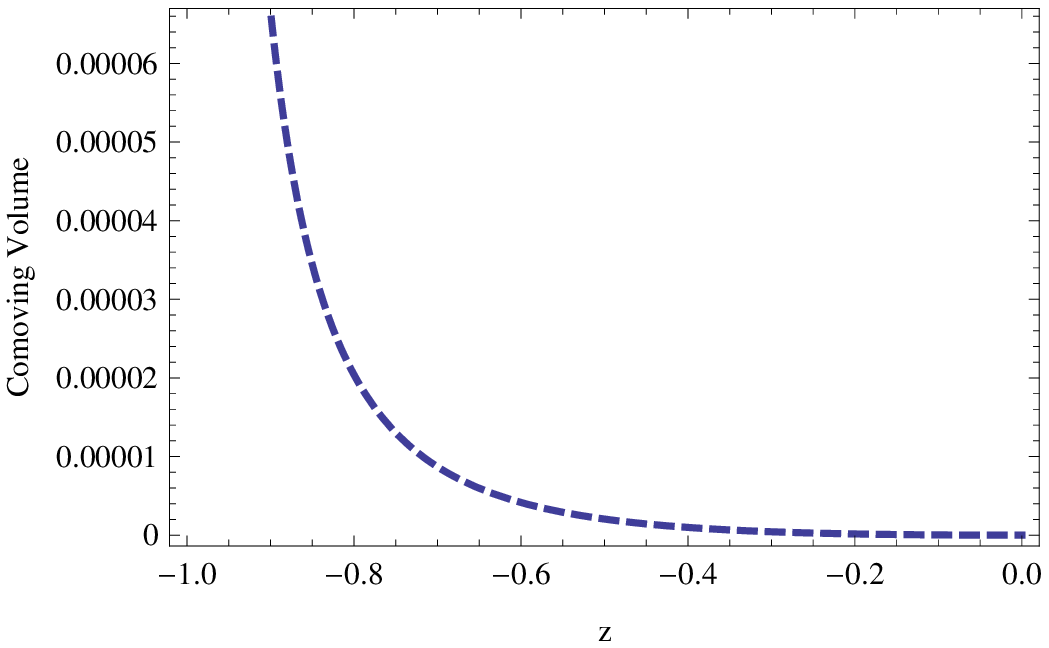}\\

~~FIG.11~~~~~~~~~~~~~~~~~~~~~~~~~~~~~~~~~~~~~~~~~~~~~~~~~~~~~~~~~~~~~~~~~~~~FIG.12\\

\caption{Comoving volume element $dV_C/dz$ as a function of
redshift assuming $H_0=72 km/s/Mpc$, $d \Omega=1$, before the
present age of the universe.}

\caption{Comoving volume element $dV_C/dz$ as a function of
redshift assuming $H_0=72 km/s/Mpc$, $d \Omega=1$, after the
present age of the universe.\\\\}
\end{figure}

\vspace{.5in}
\begin{figure}[!h]

\epsfxsize = 3.0 in \epsfysize =2 in
\epsfbox{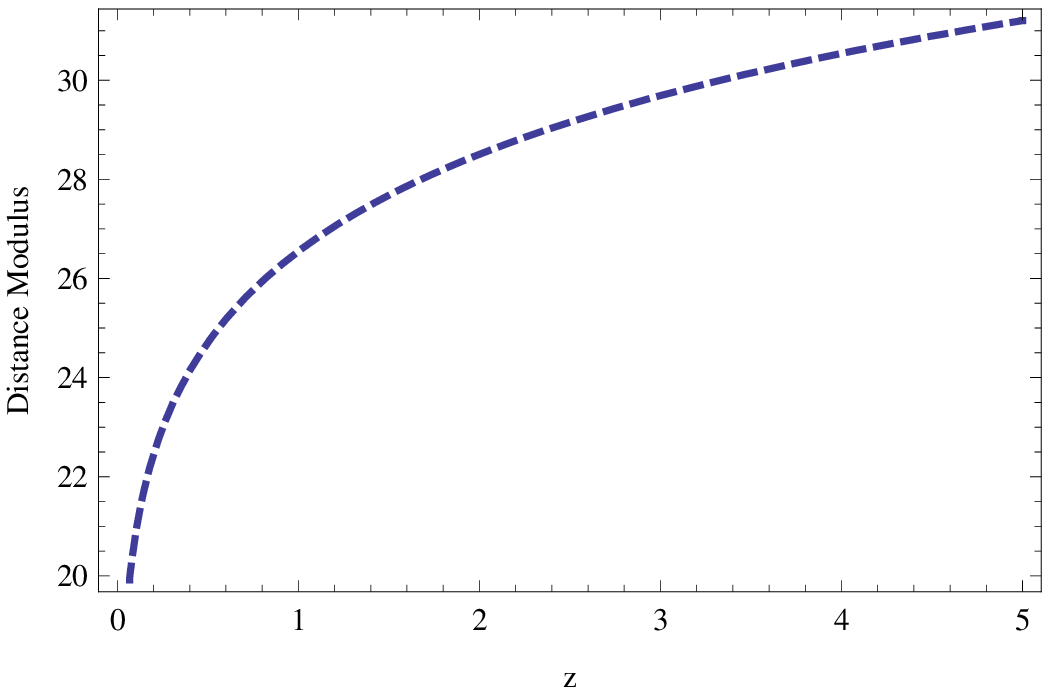}~~~~\epsfxsize = 3.0 in \epsfysize =2 in
\epsfbox{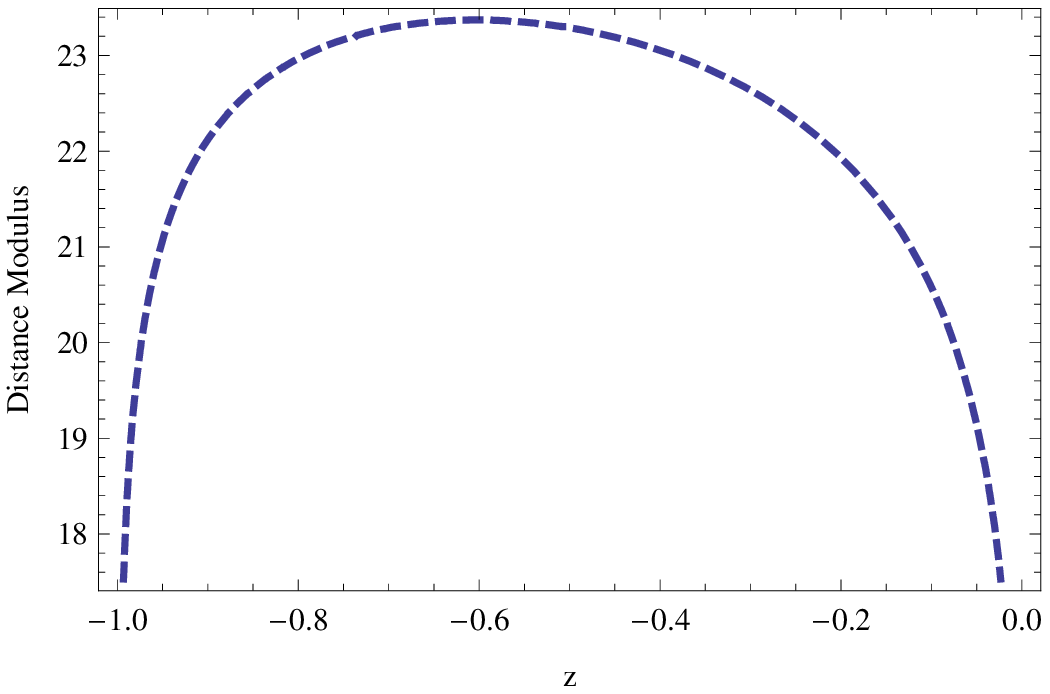}\\

~~FIG.13~~~~~~~~~~~~~~~~~~~~~~~~~~~~~~~~~~~~~~~~~~~~~~~~~~~~~~~~~~~~~~~~~~~~FIG.14\\

\caption{Distance modulus $D_M$ as a function of redshift before
the present age of the universe.}

\caption{Distance modulus $D_M$ as a function of redshift after
the present age of the universe.\\}

\end{figure}

\subsection{Distance Modulus}

The {\it distance modulus} is define by
\begin{eqnarray} D_M=5 \log \left( \frac{d_L}{10 pc}\right)\end{eqnarray}

because it is the magnitude difference between an object's
observed bolometric (i.e., integrated over all frequencies) flux
and what it would be if it were at 10 pc (this was once thought to
be the distance to Vega) and $d_L$ is the luminosity distance.
Distance modulus $D_M$ as a function of redshift before and after
the present age of the universe have been shown in figures 13 and
14. We see that for $z>0$, $D_M$ decreases as $z$ decreases but
for $z<0$, $D_M$ increases as $z$ decreases upto a certain stage
(about $z\sim -0.6$) and after that $D_M$ decreases as $z$
decreases.

\subsection{Probability of intersecting objects}

The incremental probability $dP$ that a line of sight will
intersect one of the objects in redshift interval $dz$ at redshift
$z$ is given by \cite{Hogg, Peebles}

\begin{eqnarray} dP=n(z) \sigma (z) D_H \frac{(1+z)^2}{E(z)}dz \end{eqnarray}

where $n(z)$ is the comoving number density and $\sigma (z)$ areal
cross-section. Assuming $n(z) \sigma (z)=1$, we obtain

\begin{eqnarray} dP=\frac{1}{H_0} (1+z)^{2-\frac{3X}{2}} dz\end{eqnarray}

For our model the expression of Probability of intersecting objects becomes

\begin{eqnarray} P=\frac{2}{H_0 (6-3X)}\left\{ (1+z)^{3-\frac{3X}{2}}-1 \right\}\end{eqnarray}

In figures 15 and 16 we draw intersection probability $P$ as a
function of redshift before and after the present age of the
universe respectively. We see that for $z>0$, $P$ decreases as $z$
decreases and for $z<0$, $P$ increases as $z$ decreases.

\vspace{.5in}
\begin{figure}[!h]

\epsfxsize = 3.0 in \epsfysize =2 in
\epsfbox{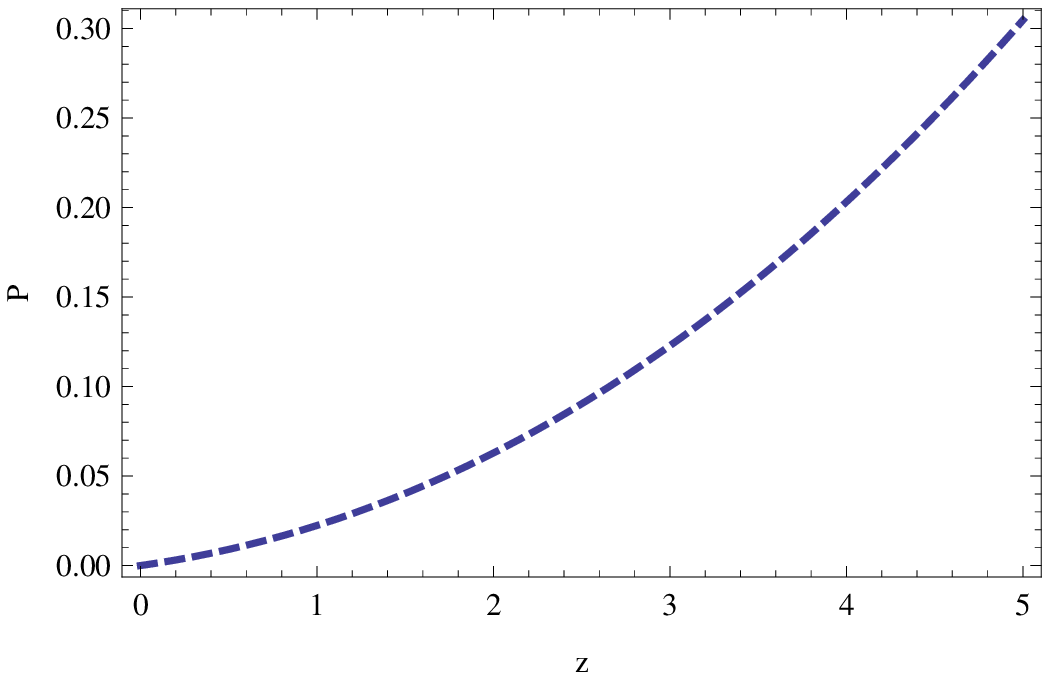}~~~~\epsfxsize = 3.0 in \epsfysize =2 in
\epsfbox{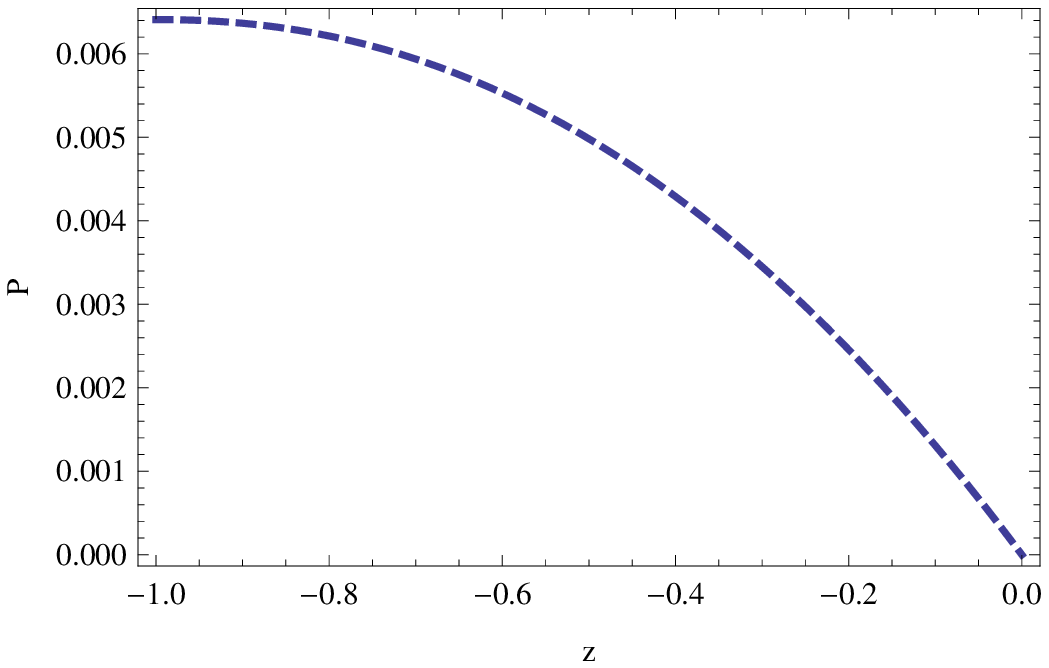}\\

~~FIG.15~~~~~~~~~~~~~~~~~~~~~~~~~~~~~~~~~~~~~~~~~~~~~~~~~~~~~~~~~~~~~~~~~~~~FIG.16\\

\caption{Intersection probability $P$ as a
function of redshift before the present age of the universe.}

\caption{Intersection probability $P$ as a function of redshift
after the present age of the universe.}
\end{figure}

\section{Discussions}

In this work, we have considered flat FRW model of the universe in
Cosmology. We have assumed that the dark energy can be considered
in the form of variable modified Chaplygin gas (VMCG). The
interaction between dark matter and VMCG has been investigated in
this model. To analyze the dynamical system, we have converted the
physical parameters into dimensionless form. We have made a
comprehensive phase-plane analysis for the VMCG model. Our main
aim was to investigate the properties of critical point of the
dynamical system which play crucially important roles for this
interacting model. We have found only one critical point which is
physically justified. Next we consider small fluctuation about the
critical point $(u_{crit},v_{crit})$ and found the eigen values of
the corresponding Jacobian matrix for $\alpha=1$. It has been
observed that the all eigen values are negative for all physical
choices of parameters and a stable attractor scaling solution is obtained.\\

In figure $1$, we plot dimensionless parameters $u$ and $v$ with
respect to $x=\ln a$. We notice that $u$ converges toward $u=1$ in
the region $0< n< 4$ (in particular, for $n=1,2,3$), also $v$
decreases simultaneously and keeps negative sign in near future.
This signifies the dark energy dominating nature of the universe
which means the energy flows from all dark matter to dark energy.
Also if we change the values of $\alpha$, then it may be seen that
nature of the dimensionless quantities $u$ and $v$ do not
sensitively depend on the change of $\alpha$. Figure 2 depicts
that the ratio $v/u=w_{vmcg}$ gets the same nature with the
variation of $n$. Since in a spatially flat universe, the
physically meaningful range of $u_{crit}$ is $0 \leq u_{crit} \leq
1$ leads to the condition $0<c\leq\frac{3\left(1+w_{dm}\right)
\left(1+\alpha\right)-n}{3 \left(1+\alpha\right)}$ along with
$n\leq min\{3(1+w_{dm})(1+\alpha),4\}$, for existence of the
critical point. This condition of $c$ represents there is an
energy transfer from to dark matter to VMCG. The phase diagram in
$u$-$v$ space (figures 3,4) shows the attractor solution for
$n=1,3$. In the above range, the critical point
$(u_{crit},v_{crit})$ is stable. Thus the present state and the
future evolution do not depend sensitively on the choice of the
initial condition. Our model cannot cross phantom divide, which
is consistent with the previous work \cite{Debnath1}, since we have
chosen power law form of $B(a)$. If we choose another form of $B(a)$,
then the VMCG model may cross the phantom barrier, which is more difficult
than this one. We may consider this type of problem in near future.\\

Moreover the expansion of the universe is governed by a power law
form around the critical point. Hence the expansion will go on
forever with an ever increasing rate. We have also studied the
evolution of the deceleration parameter $q(x) =-\frac{\ddot{a}}{a
\dot{a}^2}=-1+\frac{3}{2}\left(1+v(x)+w_{dm}(1-u(x))\right)$ for
the interacting VMCG model with dark matter. The result is shown
in figure 5 with different parameters value ranging from $n=1$ to
$n=3$. Therefore, one can observed that for $n=1,2,3$ the values
of $q$ decrease from positive value to negative value but cannot
reached $-1$. On the other way, figure 6 shows that $q$ decreases
from positive value to $-1$ for $n=0$. So VMCG ($n\ne 0$)
generates quiessence scenario for late stage but only MCG ($n=0$)
generates the $\Lambda$CDM model for late stage of the Universe.
Also we have obtained the statefinder parameters at the critical
point. Around the critical point, the natures of the parameters
are drawn in figures 7 and 8. In all the above figures, we have
chosen $c=0.01, A=0.3, w_{dm}=0.01$ and $\alpha=0.8$.\\

Distance measurements of the Universe around the critical point
have been discussed. The look back time, proper distance,
luminosity distance, angular diameter distance, comoving volume,
distance modulus and probability of intersecting objects before
and after the present age of the universe have been calculated in
terms of $z$, $X$ and $H_{0}$ and their progressions are shown in
figures 9 -16. For all figures, we have assumed $H_0=72km/s/Mpc$.
The maximum angular diameter distance has been found for a
particular value of redshift $z$. Comoving volume element
$dV_C/dz$ before and after the present age of the universe are
respectively drawn in figures 11 and 12. We see that when $z>0$,
the volume element decreases as $z$ decreases and when $z<0$, the
volume element increases as $z$ decreases. Distance modulus $D_M$
as a function of redshift before and after the present age of the
universe have been shown in figures 13 and 14. We see that for
$z>0$, $D_M$ decreases as $z$ decreases but for $z<0$, $D_M$
increases as $z$ decreases upto a certain stage (about $z\sim
-0.6$) and after that $D_M$ decreases as $z$ decreases. In figures
15 and 16 we draw intersection probability $P$ as a function of
redshift before and after the present age of the universe
respectively. We see that for $z>0$, $P$ decreases as $z$
decreases and for $z<0$, $P$ increases as $z$ decreases. So these
are the main consequences of the power law form of scale factor.
\\\\

{\bf Acknowledgement:}\\

One of the authors (JB) is thankful to CSIR, Govt of India for providing Junior Research Fellowship.\\

\end{document}